\documentclass[conference]{IEEEtran}
\IEEEoverridecommandlockouts

\usepackage{amsmath,amssymb,amsfonts}
\usepackage{dsfont} 
\usepackage{algorithmic}
\usepackage{graphicx}
\graphicspath{{./figures_comm/}{../figures_comm/}}
\usepackage{textcomp}
\usepackage{xcolor}
\usepackage{booktabs}
\usepackage{cite}
\usepackage{url}
\usepackage{siunitx}
\usepackage{epstopdf}
\usepackage[font=footnotesize,labelfont=bf,labelsep=period,skip=7pt]{caption}
\usepackage{subcaption}
\usepackage{multirow}
\usepackage{array}
\usepackage[letterpaper, 
            left=0.65in,   
            right=0.65in,  
            top=0.75in,    
            bottom=1.05in] 
            {geometry}

\AtBeginDocument{%
  \setlength{\abovedisplayskip}{6pt plus 2pt minus 3pt}%
  \setlength{\belowdisplayskip}{6pt plus 2pt minus 3pt}%
  \setlength{\abovedisplayshortskip}{2pt plus 1pt minus 1pt}%
  \setlength{\belowdisplayshortskip}{3pt plus 1pt minus 2pt}}

\newcommand{\vct}[1]{\ensuremath{\boldsymbol{#1}}}
\newcommand{\PL}{\ensuremath{\mathrm{PL}}}
\newcommand{\snr}{\ensuremath{\mathrm{SNR}}}
\newcommand{\fc}{\ensuremath{f_c}}

\DeclareSIUnit{\dB}{dB}
\DeclareSIUnit{\dBm}{dBm}
\DeclareSIUnit{\dBi}{dBi}

\title{Quantifying the Reality Gap for RL-Based UAV Placement at mmWave and Sub-THz}

\author{%
  \IEEEauthorblockN{Abdullateef Almohamad$^{1}$, Mostafa Ibrahim$^{1}$, Sabit Ekin$^{1}$, Khalid Qaraqe$^{2}$}
  \IEEEauthorblockA{$^{1}$Department of Electrical \& Computer Engineering,
                    Texas A\&M University, College Station, TX, USA \\
                    $^{2}$College of Science and Engineering, Hamad Bin Khalifa University, Doha, Qatar\\
                    Email: abdullateef.almhd@tamu.edu}
}

\begin{document}
\maketitle

\begin{abstract}
Reinforcement learning (RL) policies for unmanned aerial vehicle (UAV)
placement in mmWave and sub-terahertz networks are typically trained on
simplified analytical channels. We quantify the resulting sim-to-real gap on a real urban
map of Doha, Qatar, at carriers $\{28, 140, 183, 300\}$~GHz and altitudes
$\{50, 75, 100, 125\}$~m, evaluating three channel pipelines: an analytical
model (FSPL $+$ atmospheric absorption $+$ cuboid LoS), full
Monte-Carlo ray tracing in Sionna~RT with ITU-R~P.676-13 absorption, and a
deterministic-LoS hybrid that reuses Sionna's mesh under a closed-form
path-gain expression. We formalize the gap on the spatial SNR distribution via four
metrics, namely bias, RMSE, Jensen--Shannon divergence, and optimum-deployment
displacement. Three findings emerge: at
$28$/$140$~GHz, $\sim 70\%$ of the apparent $-5.6$/$-4.8$~dB Sionna bias is
Monte-Carlo undersampling and shrinks to $-1.7$/$-1.5$~dB after mitigation;
at $183$~GHz a $-9.2$~dB residual isolates the atmospheric absorption--ITU-R~P.676
line-shape disagreement; at $300$~GHz the stochastic ray tracer agrees
with the analytical model only coincidentally, with a $+3.8$~dB structural
offset exposed by the deterministic-LoS pipeline. Across all carriers the
linear-domain regret of the analytical-trained policy stays $\geq 0.93$,
indicating practical near-optimality but with a carrier-resolved SNR bias that warrants explicit reporting.
\end{abstract}


\section{Introduction}\label{sec:intro}

Space--air--ground (SAG) integrated networks are a core architectural
element of the $6$G vision, with unmanned aerial vehicles (UAVs) and
high-altitude platforms positioned as agile relays between dense low-Earth-orbit
constellations and ground users \cite{mozaffari2019tutorial}.
As $6$G services move beyond millimetre-wave
(mmWave, \SIrange{30}{100}{\giga\hertz}) into the sub-terahertz (sub-THz,
\SIrange{100}{300}{\giga\hertz}) regime
\cite{rappaport2019wireless,akyildiz2014terahertz}, the propagation channel
becomes both highly directional and strongly carrier-selective. The same urban
geometry that yields broad quasi-omnidirectional coverage at $\SI{28}{\giga\hertz}$
collapses into a thin line-of-sight (LoS) service shell at
$\SI{300}{\giga\hertz}$, where atmospheric absorption can reach
\SIrange{10}{50}{\decibel\per\kilo\meter} \cite{itur676} and dominates the
link budget. Where to place the relay is therefore a per-carrier
optimization problem with substantial topographic structure.

Closed-form altitude and 3-D placement analyses under probabilistic LoS
\cite{alhourani2014optimal,kalantari20163d,mozaffari2019tutorial}, and
convex relaxations for multi-UAV cases \cite{lyu2017placement}, predate the
learning-based approach but typically assume frequency-flat channels and do
not extend cleanly to sub-THz, where absorption is strongly
frequency-selective and structured by atmospheric resonances.
Reinforcement learning (RL) has since emerged as a flexible tool for UAV
placement and trajectory \cite{bayerlein2018trajectory,liu2019trajectory},
but training requires channel evaluations at volumes that real measurements
or full-wave ray tracing cannot supply. The customary trade-off is to train
on a simplified analytical channel and validate the learned policy on a higher-fidelity reference; most prior RL works train and
evaluate within the same simulator, so any disagreement with a
physics-grade channel is silent. This disagreement is the wireless analogue
of the well-studied sim-to-real or reality gap in robotics
\cite{tobin2017domain,peng2018simtoreal}: when the optima of the training
and validation channels disagree, the policy fails to transfer.

Three observations motivate this work. First,
quantitative reality-gap studies for THz UAV networks are scarce. Existing
work either compares analytical models against one another, restricts ray
tracing to a fixed transmit position, or considers a single carrier.
Second, the open release of Sionna~RT~\cite{hoydis2023sionna}, a
GPU-accelerated, differentiable Monte-Carlo ray tracer with ITU-R~P.676-13
atmospheric absorption \cite{itur676}, now makes large-scale carrier-aware
urban ray tracing tractable, allowing the same RL evaluation to be replayed
on a physics-grade channel. Third, the simplified
$100$--$450$~\unit{\giga\hertz} Kokkoniemi--Lehtom\"aki absorption fit
\cite{kokkoniemi2021los} commonly used in analytical THz models, six
Lorentzian resonances plus a polynomial water-vapour continuum, is
accurate to within a few~\unit{\dB} per kilometre away from strong lines
but can deviate materially from ITU-R~P.676-13 in the immediate
neighborhood of the $183$~\unit{\giga\hertz} water-vapor and
$325$~\unit{\giga\hertz} oxygen resonances; furthermore, the standard
Monte-Carlo ray-tracing pipeline introduces a non-trivial sampling
artifact that, if uncorrected, masquerades as a physical disagreement and
inflates the apparent gap by an order of magnitude.

\begin{figure*}[!t]
\centering
\begin{subfigure}[b]{0.28\linewidth}
  \centering
  \includegraphics[width=\linewidth,trim=4mm 20mm 45mm 20mm,clip]{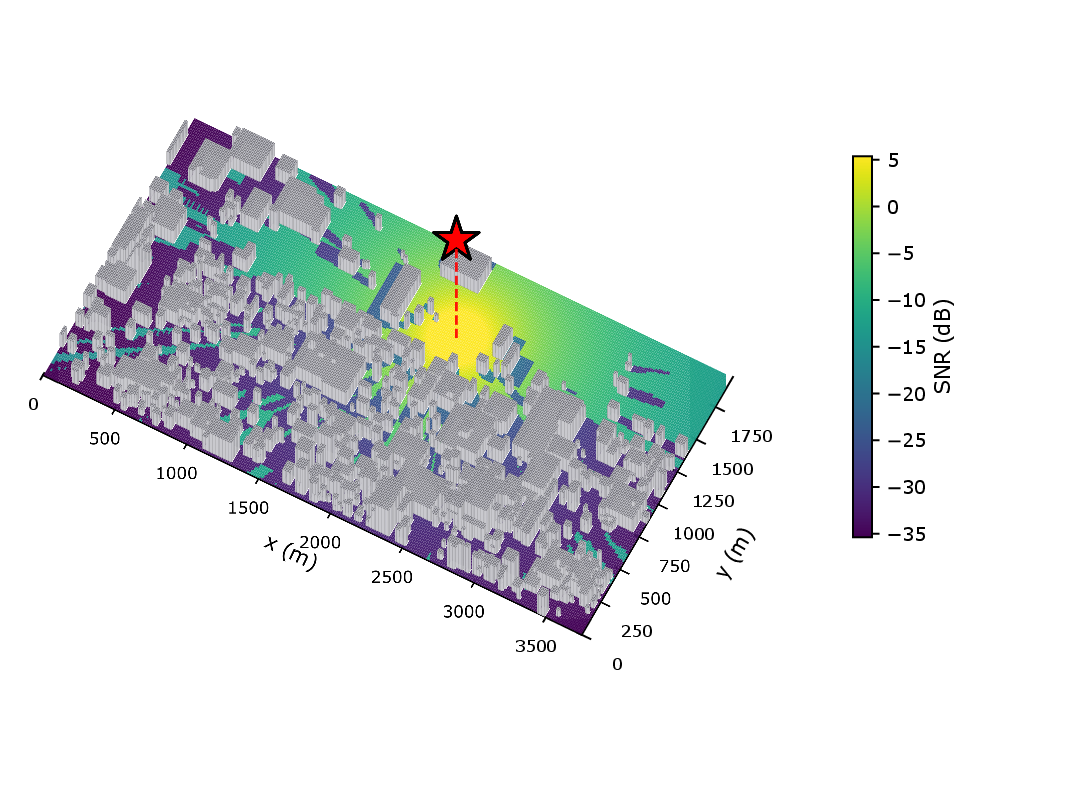}
  \caption{Analytical $\mathcal{M}$ (FSPL $+$ absorption loss $+$ merged-cuboid LoS).}
  \label{fig:cov3d-M}
\end{subfigure}\hfill
\begin{subfigure}[b]{0.28\linewidth}
  \centering
  \includegraphics[width=\linewidth,trim=1mm 4mm 30mm 20mm,clip]{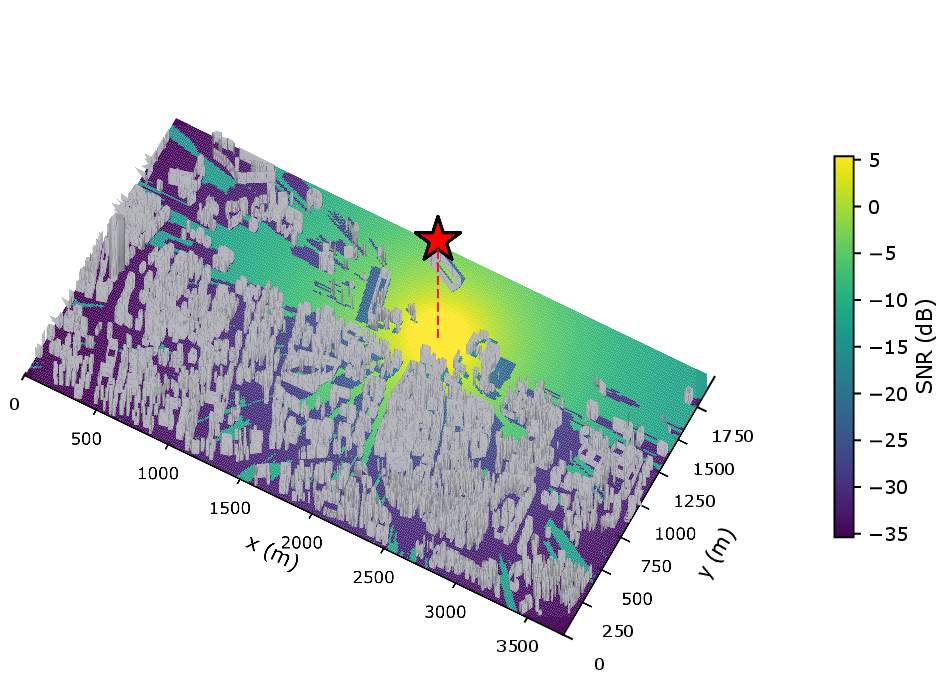}
  \caption{Deterministic-LoS $\mathcal{L}$ (Sionna mesh $+$ FSPL $+$ ITU-R~P.676-13).}
  \label{fig:cov3d-L}
\end{subfigure}\hfill
\begin{subfigure}[b]{0.28\linewidth}
  \centering
  \includegraphics[width=\linewidth,trim=100mm 20mm 45mm 20mm,clip]{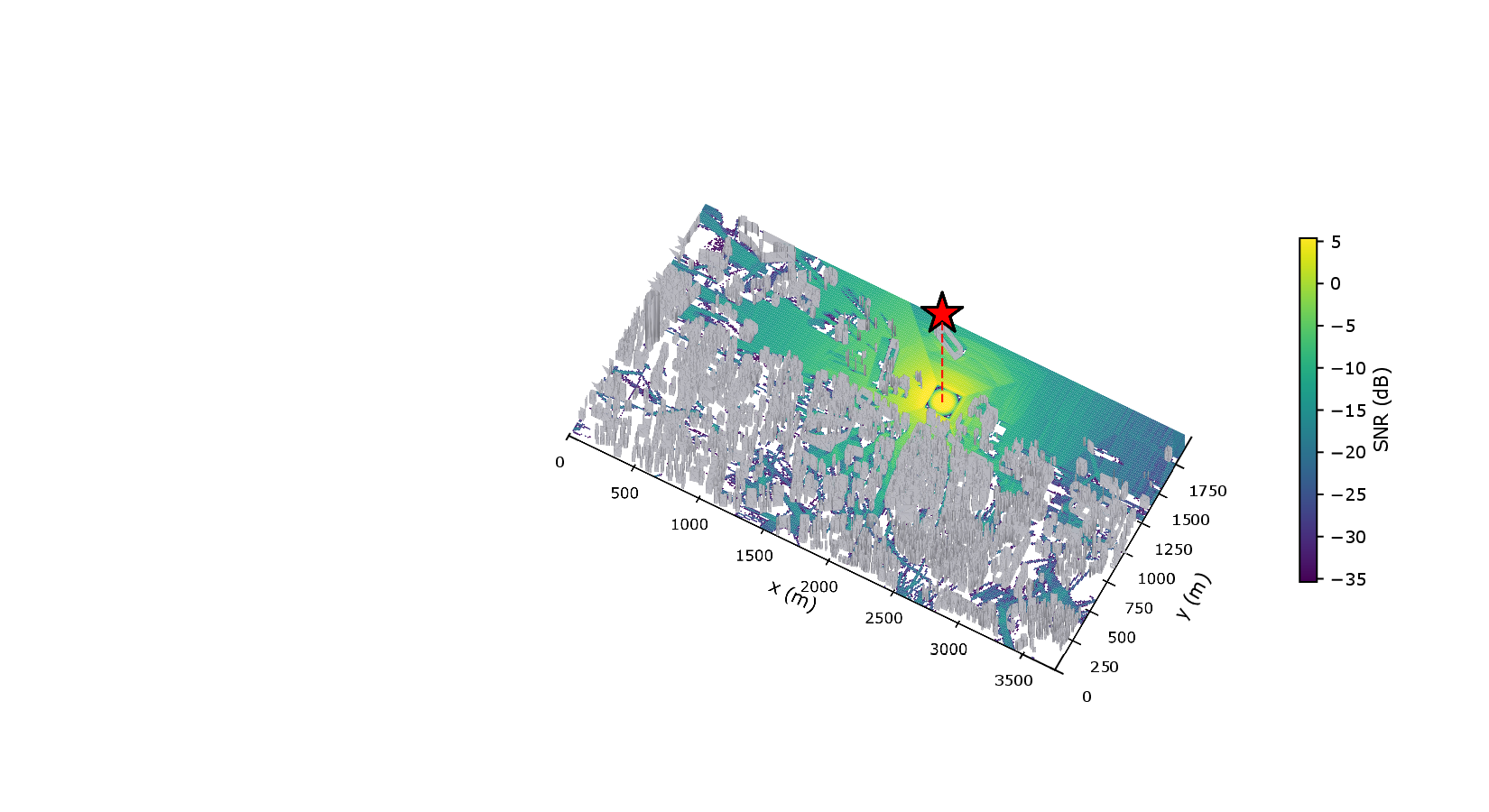}
  \caption{Stochastic ray-traced $\mathcal{S}$ (Sionna RT, $D{=}1$ specular reflections, ITU-R~P.676-13).}
  \label{fig:cov3d-S}
\end{subfigure}
\caption{Ground-plane SNR coverage at $z = \SI{100}{\meter}$, $\fc = \SI{28}{\giga\hertz}$, rendered against each pipeline's building geometry and physics.}
\label{fig:cov3d-3way}
\end{figure*}
Our objective is to deliver a quantitative,
per-carrier characterization of the reality gap incurred by an
analytically-trained RL UAV-placement policy when deployed against
ray-traced channels, and to disentangle Monte-Carlo sampling artifacts
from genuine physics disagreement so that practitioners know which component of the simulation stack to upgrade for a given
carrier. To this end, this paper makes four contributions.
\emph{(i)} A reproducible reality-gap benchmark for RL-driven UAV
placement over a real urban map, at four carriers
\{$28, 140, 183, 300$\}~\unit{\giga\hertz} and four altitudes
\{$50, 75, 100, 125$\}~\unit{\meter}.
\emph{(ii)} a position-distributed reality-gap framework based on
the per-deployment SNR distribution, with four operational
metrics, namely bias, RMSE, Jensen--Shannon (JS) divergence, and optimum-deployment
displacement, that decouple systematic offset, distributional spread, and
decision-level error, complemented by a linear-domain regret ratio that
quantifies the policy-transfer cost directly.
\emph{(iii)} a deterministic-LoS pipeline that
reuses Sionna's mesh geometry under a closed-form path-gain expression,
enabling an exact decomposition of the apparent Monte-Carlo gap into a
sampling-induced and a physics-induced component.
\emph{(iv)} a characterization of when a
policy trained on the analytical channel transfers to ray tracing, 
sampling-dominated at $28$/$140$~\unit{\giga\hertz}, absorption-model
dominated at $183$~\unit{\giga\hertz}, and structurally different despite a
coincidentally small bias at $300$~\unit{\giga\hertz}.

The remainder of the paper is organized as follows.
Section~\ref{sec:system} formalizes the three channel pipelines and the
placement MDP; Section~\ref{sec:method} describes the RL training, Sionna
processing, and the reality-gap framework; Section~\ref{sec:results}
reports numerical results; and Section~\ref{sec:conclusion} concludes.

\section{System Model}\label{sec:system}
A single rotary-wing UAV hovers at a fixed altitude $z$ over a dense
urban area and provides broadband downlink coverage to ground users uniformly distributed across the served area. Isotropic single-antenna terminals are assumed on both ends. The latter two assumptions can be relaxed for an extended version of this study.

The UAV operates at a single carrier
$\fc \in \{28, 140, 183, 300\}$~\unit{\giga\hertz}, spanning mmWave to
sub-THz, and at one of the altitudes
$z \in \{50, 75, 100, 125\}$~\unit{\meter}. Because blockage and absorption are strongly carrier-dependent, the optimal hover position varies with $\fc$, so a multi-band UAV would
still solve four independent placement problems which we treat with a dedicated RL agent each (\S\ref{sec:mdp}). The operator's
task is to choose the hover position $\vct p_u = (u_x, u_y, z)$ that
maximizes the aggregate signal quality delivered to the
ground-user population, where aggregating over users is the fairness-agnostic coverage objective
standard in UAV placement. The remainder of this section formalizes the
geometry, the three channel pipelines, and the per-position SNR distribution that drives both the RL reward and the reality-gap analysis.

\subsection{Geometry and Building Model}\label{sec:scenario}

The served area is a $\SI{3.75}{\kilo\meter}\times\SI{1.98}{\kilo\meter}$
section of central Doha, Qatar, discretized on a
uniform Cartesian grid
$\mathcal{G} = \{(x,y) : 0 \le x < G_x,\, 0 \le y < G_y\}$ with $G_x=375$,
$G_y=198$, and cell size $\Delta = \SI{10}{\meter}$. The UAV operates at
one of four altitudes
$\mathcal{Z} = \{50, 75, 100, 125\}$~\unit{\meter}, indexed by $z$. Building
geometry is extracted from a triangulated 3-D mesh (STL) and pre-processed into axis-aligned cuboids whose footprints are
iteratively merged when contiguous and of comparable height (within
$\SI{1}{\meter}$) to reduce the number of buildings and computation cost; the resulting per-cell height map
$h: \mathcal{G} \to \mathbb{R}_{\geq 0}$ admits only positions with
$h(x,y) < z$ as deployable, defining the safe-deployment set
\begin{equation}\label{eq:safeset}
\mathcal{P}_z = \big\{\vct p_u = (u_x, u_y, z) : h(u_x, u_y) < z \big\},
\end{equation}
Carriers are indexed by $f \in \{0,1,2,3\}$ corresponding to
$\fc \in \{28, 140, 183, 300\}$~\unit{\giga\hertz}.

\subsection{Three Channel Pipelines}\label{sec:channels}

For UAV position $\vct p_u$ and ground cell $\vct p_r = (r_x, r_y, 0)$ with
link distance $d \triangleq \|\vct p_u - \vct p_r\|_2$, we evaluate three
channel pipelines, namely analytical ($\mathcal{M}$), Sionna stochastic RT ($\mathcal{S}$) and Sionna deterministic ($\mathcal{L}$), each producing a linear path gain
$g_{i}(\vct p_u, \vct p_r, \fc)$ for
$i \in \mathcal{I} \triangleq \{\mathcal{M}, \mathcal{S}, \mathcal{L}\}$.

\subsubsection{Analytical pipeline ($\mathcal{M}$)}\label{sec:M}
The analytical pipeline expresses the link gain as a product of three factors, free-space path loss (FSPL), molecular absorption, and a
geometric LoS indicator:
\begin{equation}\label{eq:matlab}
\begin{aligned}
g_{\mathrm{M}}(\vct p_u, \vct p_r, \fc)
   &= \underbrace{\Big(\frac{c}{4\pi \fc d}\Big)^{\!2}}_{\text{FSPL}}
   \cdot \underbrace{e^{-\tfrac{1}{2}\, d\, \Sigma_{\mathrm{K}}(\fc)}}_{\text{absorption}} \\
   &\quad \cdot \underbrace{\mathds{1}\!\big[\,\mathsf{LoS}_{\mathrm{cube}}(\vct p_u, \vct p_r)\,\big]}_{\text{LoS}},
\end{aligned}
\end{equation}
where $c$ is the speed of light. The absorption coefficient
$\Sigma_{\mathrm{K}}(\fc)$ is the simplified $100$--$450$~\unit{\giga\hertz}
Kokkoniemi--Lehtom\"aki molecular-absorption fit \cite{kokkoniemi2021los},
comprising six Lorentzian resonance lines $183.3$, $325.2$, $380.4$, $439.5$, and
$448.2$~\unit{\giga\hertz}, and a polynomial water-vapor continuum,
evaluated at the standard atmosphere $T=\SI{293.15}{\kelvin}$,
$p=\SI{1013}{\hecto\pascal}$, and relative humidity $\phi=0.5$. The LoS
predicate $\mathsf{LoS}_{\mathrm{cube}}$ is a 2.5-D rooftop-projection
blockage test against a merged axis-aligned cuboid building representation;
non-LoS cells receive zero linear gain, i.e. no reflections or scattering.

\subsubsection{Stochastic ray-traced pipeline ($\mathcal{S}$)}\label{sec:S}
The stochastic pipeline replaces the closed-form simplicity of $\mathcal{M}$
with a physics-grade evaluation of $g_{\mathrm{S}}$ on the actual mesh
geometry. Sionna~RT~\cite{hoydis2023sionna} performs Monte-Carlo path
tracing with $N_s = 10^{5}$ rays per UAV position, maximum bounce depth
$D = 1$, and specular reflection only (no diffraction or scattering), on a
closed-mesh triangular building representation derived from the same
dataset; the omission of diffraction and diffuse scattering understates
non-LoS energy near building edges. Atmospheric absorption follows ITU-R~P.676-13~\cite{itur676}
parameterized by ($T$, $p$, water-vapour density
$\rho_w = \SI{12}{\gram\per\meter\cubed}$); material electrical properties
$(\varepsilon_r, \sigma)$ are taken from the ITU-R~P.2040
database~\cite{itur2040}. Atmospheric attenuation is applied per
ground-cell distance after the ray-tracing call.

\subsubsection{Deterministic-LoS pipeline ($\mathcal{L}$)}\label{sec:L}
To isolate sampling-induced effects from genuine modeling differences, we
introduce a third pipeline that re-uses Sionna's mesh geometry under a
deterministic LoS test. The path gain is computed in closed form as
\begin{equation}\label{eq:los}
g_{\mathrm{L}}(\vct p_u, \vct p_r, \fc)
   = \Big(\frac{c}{4\pi \fc d}\Big)^{\!2}
     \,e^{-\tfrac{1}{2}\, d\, \gamma_{\mathrm{ITU}}(\fc)}\,
     \mathds{1}\!\big[\,\mathsf{LoS}_{\mathrm{mesh}}(\vct p_u, \vct p_r)\,\big],
\end{equation}
where $\mathsf{LoS}_{\mathrm{mesh}}$ is the deterministic ray-AABB (Axis-Aligned Bounding Box)
occlusion test implemented in Mitsuba \cite{NimierDavidVicini2019Mitsuba2}, and $\gamma_{\mathrm{ITU}}(\fc)$ is
the ITU-R~P.676-13 specific attenuation coefficient. By construction,
$\mathcal{L}$ shares geometry and absorption physics with $\mathcal{S}$ but
exposes neither Monte-Carlo variance nor multi-bounce reflections; it is
therefore a clean ablation point that decouples the two sources of disagreement between $\mathcal{M}$ and $\mathcal{S}$, namely geometry/physics and sampling effects.

\subsubsection{Per-position aggregate path loss}\label{sec:plavg}
For each pipeline $i \in \mathcal{I}$ the
aggregate path loss at UAV position $\vct p_u$ is
\begin{equation}\label{eq:plavg}
\PL^{i}(\vct p_u, \fc)
   = -10 \log_{10}\!\Bigg(\frac{1}{|\mathcal{G}|}
        \sum_{\vct p_r \in \mathcal{G}}
        g_{i}(\vct p_u, \vct p_r, \fc)\Bigg) \;[\unit{\dB}],
\end{equation}
which describes the mean linear gain provided by the UAV to a uniformly
distributed ground-user population. Because \eqref{eq:plavg} is invariant
to transmit power, the three pipelines are calibrated identically and any
disagreement reflects channel modeling rather than link-budget choices.

\subsection{Per-Position SNR and Deployment Distribution}\label{sec:snr-dist}

Given a generic carrier-dependent link budget $(P_t, G_t, G_r, \eta, B(\fc))$,
the per-position receive SNR under pipeline $i \in \mathcal{I}$ is
\begin{equation}\label{eq:snr}
\snr^{i}(\vct p_u, \fc)
   = P_t + G_t + G_r - \PL^{i}(\vct p_u, \fc) - \mathcal{N}_0(\fc),
\end{equation}
with thermal noise floor
$\mathcal{N}_0(\fc) = -174 + 10\log_{10} B(\fc) + \eta$~\unit{\dBm}.
Treating the UAV position as a uniform random variable over $\mathcal{P}_z$,
we obtain the empirical \emph{position-distributed} SNR
\begin{equation}\label{eq:phi}
\Phi_{z,f}^{i}
  \,\triangleq\, \big\{\snr^{i}(\vct p_u, \fc) : \vct p_u \in \mathcal{P}_z\big\},
\end{equation}
which is the central object of our reality-gap analysis.

\subsection{UAV Deployment as a Markov Decision Process}\label{sec:mdp}

We cast deployment as a finite-horizon MDP
$\big\langle \mathcal{X}, \mathcal{A}, P, r, \gamma_{\mathrm{rl}} \big\rangle$
on the discretised grid. The state $\vct x_t \in [0,1]^{9}$ encodes
normalised UAV position, four-direction occlusion ray-cast distances, the
local signal quality, and the time remaining; the action set
$\mathcal{A} = \{\mathrm{hover}, \pm x, \pm y, \pm z\}$ has cardinality
$7$; transitions $P$ are deterministic; the discount is
$\gamma_{\mathrm{rl}} = 0.995$; and the per-step reward is
\begin{equation}\label{eq:reward}
\begin{aligned}
r_t &= \underbrace{w_s\, s_t + w_c\, c_t}_{r_{\mathrm{quality}}\,\in\,[0,1]}
   + \underbrace{\alpha\,(q_t - q_{t-1})\,\mathds{1}[a_t \neq \mathrm{hover}]}_{r_{\mathrm{shape}}}\\
   &\quad + \underbrace{\beta\, q_T\, \mathds{1}[t=T]}_{r_{\mathrm{terminal}}},
\end{aligned}
\end{equation}
with $w_s=0.6$, $w_c=0.4$, $\alpha=0.5$, $\beta=5$, episode horizon $T=500$,
and quality proxy $q_t = w_s s_t + w_c c_t$. The signal proxy
$s_t \in [0,1]$ is the min-max normalisation of $\PL^{\mathrm{M}}$ over
$\mathcal{P}_z$; the coverage proxy $c_t \in [0,1]$ is the LoS fraction over
$\mathcal{G}$; building collisions terminate the episode with reward $-10$.

\section{Methodology}\label{sec:method}

\subsection{RL Training}\label{sec:method-rl}

For each carrier we train an independent PPO agent~\cite{schulman2017ppo}
on the analytical channel of \eqref{eq:matlab} using
Stable-Baselines3~\cite{stable-baselines3}. Hyperparameters: $N_e = 8$
parallel environments, $T_{\mathrm{tot}} = 3{\times}10^{6}$ environment
steps, linearly annealed learning rate
$3{\times}10^{-4}\!\to\!1{\times}10^{-5}$, GAE $\lambda=0.95$, clip range
$0.2$, mini-batch size $512$, entropy coefficient $0.01$, and a
$2{\times}128$-tanh MLP policy. Reward is normalised by a running estimator;
observations are not (they are already in $[0,1]^{9}$). Each policy converges
in $\approx \SI{4}{\hour}$ on a single RTX-class GPU. The trained policies
are then evaluated under the three channel pipelines without further
fine-tuning.

\subsection{Sionna Pre-computation}\label{sec:method-precompute}

Calling the ray tracer at every RL step is computationally intractable. We
instead pre-compute the four-tensor
$\PL^{\mathrm{S}} \in \mathbb{R}^{4 \times 4 \times G_x \times G_y}$ on a
strided sub-grid (stride $\sigma = 5$ along $x$ and $y$) and bilinearly
interpolate to full resolution. After building height masking via
\eqref{eq:safeset}, this yields $|\mathcal{P}_z| \approx 1.6{\times}10^{4}$
deployable positions per altitude. Each Sionna call invokes
\texttt{RadioMapSolver} with the parameters of \S\ref{sec:S}; ITU-R~P.676-13
absorption is applied per ground-cell distance after the ray-tracing step,
ensuring consistent treatment across the grid. Total pre-computation time
is $\sim\SI{3}{\hour}$ per carrier on a Colab L4 GPU.

\subsection{Sampling-Induced Bias and the Deterministic-LoS Construct}\label{sec:method-los}

The stochastic Sionna pipeline marks a ground cell as covered only if at
least one ray reaches it. Since the $N_s = 10^{5}$ rays are launched over
$4\pi$ steradians, a ground cell of area $\Delta^{2}$ at link distance $d$
subtends a solid angle $\approx \Delta^{2} z / d^{3}$ and receives an
expected ray-hit count of
$\bar{n}(d) \approx N_s \Delta^{2} z \big/ 4\pi d^{3}$,
so under a Poisson approximation cells beyond the critical distance
$d_c = \big(N_s \Delta^{2} z / 4\pi\big)^{1/3}$, at which $\bar{n}=1$, are
missed with probability approaching one. At $z=\SI{100}{\meter}$,
$d_c \approx \SI{430}{\meter}$: the well-sampled disk around the UAV spans
only ${\approx}\,7\%$ of the served area, consistent with the empirical
LoS coverage of $\mathcal{S}$ of $\approx 0.06$ versus $\approx 0.70$ for
$\mathcal{M}$, and the $d_c \propto z^{1/3}$ scaling reproduces the
altitude trend of Table~\ref{tab:cov}. We stress that the
sampling/physics decomposition of \S\ref{sec:method-decomp} is measured
directly from the empirical $\mathcal{S}$--$\mathcal{L}$ comparison and
does not rely on this model. To separate this Monte-Carlo artifact from
genuine modeling differences, we evaluate the deterministic-LoS pipeline
of \eqref{eq:los}, casting one ray to every ground-cell centre, testing
occlusion against the Mitsuba scene graph, and constructing the full
$\PL^{\mathrm{L}}$ tensor in closed form from the deterministic LoS map,
FSPL, and ITU-R~P.676 absorption, with no Monte-Carlo variance at a
fraction of the stochastic compute.

\subsection{Reality-Gap Framework}\label{sec:method-gap}

Let $\Phi_{z,f}^{i}$ denote the empirical position-distributed SNR
\eqref{eq:phi} under pipeline $i \in \mathcal{I}$, and let
$\mathcal{T} \in \{\mathrm{S}, \mathrm{L}\}$ index a ray-traced target. We
quantify the disagreement between the analytical reference $\mathcal{M}$ and
$\mathcal{T}$ via four reality-gap metrics that decouple systematic offset,
distributional spread, and decision-level error: the mean bias
\eqref{eq:bias} captures the systematic dB offset; the position-wise
RMSE \eqref{eq:rmse} captures the pointwise spread around the bias; the
Jensen--Shannon divergence \eqref{eq:js} captures the population-level
distributional disagreement; and the optimum-deployment displacement
\eqref{eq:disp} captures the decision-level error between per-pipeline
argmaxes. A linear-domain regret ratio \eqref{eq:regret} converts the
displacement into an operational SNR cost.
\begin{align}
\mathrm{bias}^{\mathcal{T}}_{f,z}
   &\,\triangleq\, \mathbb{E}_{\vct p_u \sim \mathcal{U}(\mathcal{P}_z)}
       \!\big[\snr^{\mathcal{T}} - \snr^{\mathrm{M}}\big], \label{eq:bias}\\
\mathrm{RMSE}^{\mathcal{T}}_{f,z}
   &\,\triangleq\, \sqrt{\mathbb{E}_{\vct p_u}\!\big[\big(\snr^{\mathcal{T}} - \snr^{\mathrm{M}}\big)^{2}\big]}, \label{eq:rmse}\\
D_{\mathrm{JS}}^{\mathcal{T}}
   &\,\triangleq\, \tfrac{1}{2} D_{\mathrm{KL}}\!\big(\Phi^{\mathrm{M}} \,\|\, \bar{\Phi}\big)
   + \tfrac{1}{2} D_{\mathrm{KL}}\!\big(\Phi^{\mathcal{T}} \,\|\, \bar{\Phi}\big), \label{eq:js}\\
\delta^{\mathcal{T}}_{f,z}
   &\,\triangleq\, \big\| \vct p_u^{\mathcal{T}\star} - \vct p_u^{\mathrm{M}\star} \big\|_{2}, \label{eq:disp}\\
\rho^{\mathcal{T}}_{f,z}
   &\,\triangleq\, \frac{\mathrm{snr}^{\mathcal{T}}_{\mathrm{lin}}\!\big(\vct p_u^{\mathrm{M}\star}\big)}
                    {\mathrm{snr}^{\mathcal{T}}_{\mathrm{lin}}\!\big(\vct p_u^{\mathcal{T}\star}\big)} \in (0,1], \label{eq:regret}
\end{align}
with $\bar{\Phi}\!=\!(\Phi^{\mathrm{M}}\!+\!\Phi^{\mathcal{T}})/2$,
$D_{\mathrm{KL}}$ the Kullback--Leibler divergence over $80$-bin empirical
histograms with shared support ($D_{\mathrm{JS}}\in[0,\ln 2]$, in nats), and
$\vct p_u^{i\star}\!\triangleq\!\arg\max_{\vct p_u\in\mathcal{P}_z}\snr^{i}(\vct p_u,\fc)$.
The dB metrics \eqref{eq:bias}--\eqref{eq:js} and the regret
\eqref{eq:regret} are invariant to the link-budget offset
$P_t+G_t+G_r-\mathcal{N}_0(\fc)$, so comparisons across pipelines reflect
channel modelling rather than link-budget choices; $\delta^{\mathcal{T}}$
is in metres.

\subsection{Decomposition into Sampling and Physics Components}\label{sec:method-decomp}

Comparing the bias against the two targets $\{\mathrm{S}, \mathrm{L}\}$
yields a clean decomposition:
\begin{equation}\label{eq:decomp}
\underbrace{\mathrm{bias}^{\mathrm{S}}_{f,z}}_{\text{apparent gap}}
   = \underbrace{\big(\mathrm{bias}^{\mathrm{S}}_{f,z} - \mathrm{bias}^{\mathrm{L}}_{f,z}\big)}_{\substack{\text{sampling}\\\text{component}}}
   + \underbrace{\mathrm{bias}^{\mathrm{L}}_{f,z}}_{\substack{\text{physics}\\\text{component}}},
\end{equation}
in which the first term captures Monte-Carlo undersampling (since
$\mathcal{S}$ and $\mathcal{L}$ share geometry and absorption physics and
differ only in sampling and reflections), and the second term captures the
residual disagreement between $\mathcal{M}$ and a deterministic ray-traced
reference. As we show in \S\ref{sec:results}, this decomposition is the
operative diagnostic that distinguishes carriers requiring physics
upgrades from carriers requiring only sampling upgrades.

\section{Results}\label{sec:results}

\subsection{Setup}\label{sec:res-setup}

Unless stated otherwise, results are reported at altitude
$z = \SI{100}{\meter}$, where $|\mathcal{P}_z| \approx 1.39{\times}10^{4}$.
The link budget assumes $P_t = \SI{30}{\dBm}$, $G_t = G_r = \SI{0}{\dBi}$
(isotropic),
$\eta = \SI{8}{\dB}$, $T_0 = \SI{290}{\kelvin}$, and per-carrier bandwidth
$B(\fc) \in \{200, 1000, 2000, 5000\}$~\unit{\mega\hertz} for
$\fc \in \{28, 140, 183, 300\}$~\unit{\giga\hertz}. All bias, RMSE,
$D_{\mathrm{JS}}$ and $\rho$ values are link-budget-invariant; only the
absolute SNR values shift with the link-budget offset.

\subsection{LoS Coverage}\label{sec:res-cov}

Table~\ref{tab:cov} reports the mean LoS fraction over $\mathcal{P}_z$ for
the three pipelines. Two observations follow.
\emph{(i)} The stochastic Sionna LoS fraction is an order of magnitude
below the analytical model at every altitude, e.g.\
$0.057$ vs.\ $0.700$ at $z=\SI{100}{\meter}$, in line with the Monte-Carlo
undersampling analysis of \S\ref{sec:method-los}.
\emph{(ii)} The deterministic-LoS pipeline recovers a substantially larger
LoS fraction (e.g., $0.283$ at $\SI{100}{\meter}$) but does not reach the
analytical value. The remaining gap reflects a genuine geometric
difference between MATLAB's merged-cuboid building model and Sionna's
closed-mesh representation. As shown below, this geometric discrepancy
translates only to a small ($\sim 1.7$~\unit{\dB}) PL bias at
$28$~\unit{\giga\hertz} because the cells that are LoS in MATLAB but not
in Sionna sit at long distances from the UAV and contribute negligible
linear gain.

\begin{table}[t]
\caption{Mean LoS coverage over $\mathcal{P}_z$ (frequency-independent).}
\label{tab:cov}
\centering\small
\setlength{\tabcolsep}{4pt}
\begin{tabular}{lccc}
\toprule
Altitude & Analytical $\mathcal{M}$ & Sionna stoch.\ $\mathcal{S}$ & Sionna-LoS $\mathcal{L}$ \\
\midrule
$\SI{50}{m}$  & $0.679$ & $0.035$ & $0.160$ \\
$\SI{75}{m}$  & $0.703$ & $0.049$ & $0.230$ \\
$\SI{100}{m}$ & $0.700$ & $0.057$ & $0.283$ \\
$\SI{125}{m}$ & $0.712$ & $0.063$ & $0.325$ \\
\bottomrule
\end{tabular}
\end{table}

\subsection{Position-Distributed SNR}\label{sec:res-snr}

\begin{figure*}[t]
\centering
\includegraphics[trim= 0mm 0mm 0mm 6mm, clip, width=0.86\linewidth]{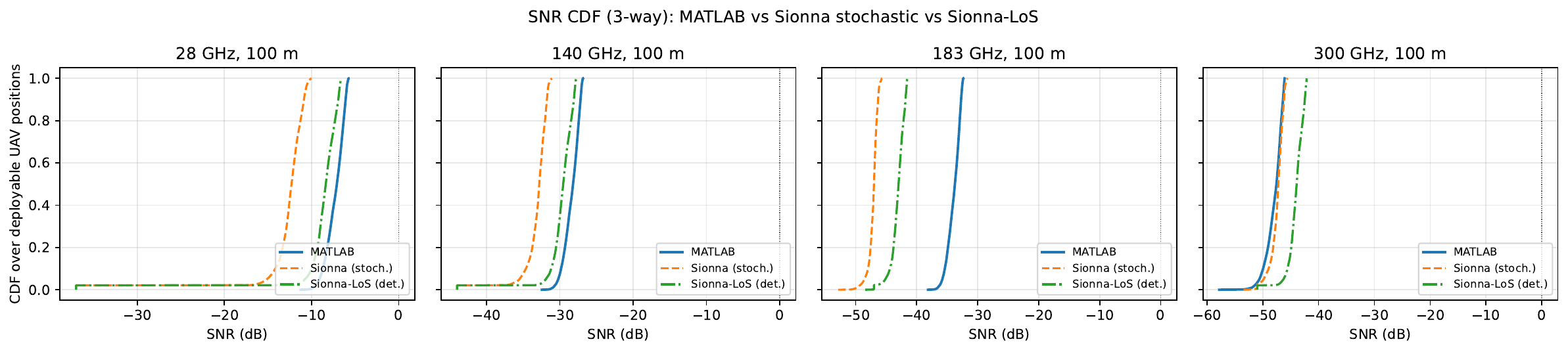}
\caption{Empirical CDFs of the position-distributed SNR
$\Phi^{\mathrm{M}}, \Phi^{\mathrm{S}}, \Phi^{\mathrm{L}}$ at
$z = \SI{100}{\meter}$.}
\label{fig:cdfs}
\end{figure*}

\begin{figure}[t]
\centering
\includegraphics[width=0.86\linewidth]{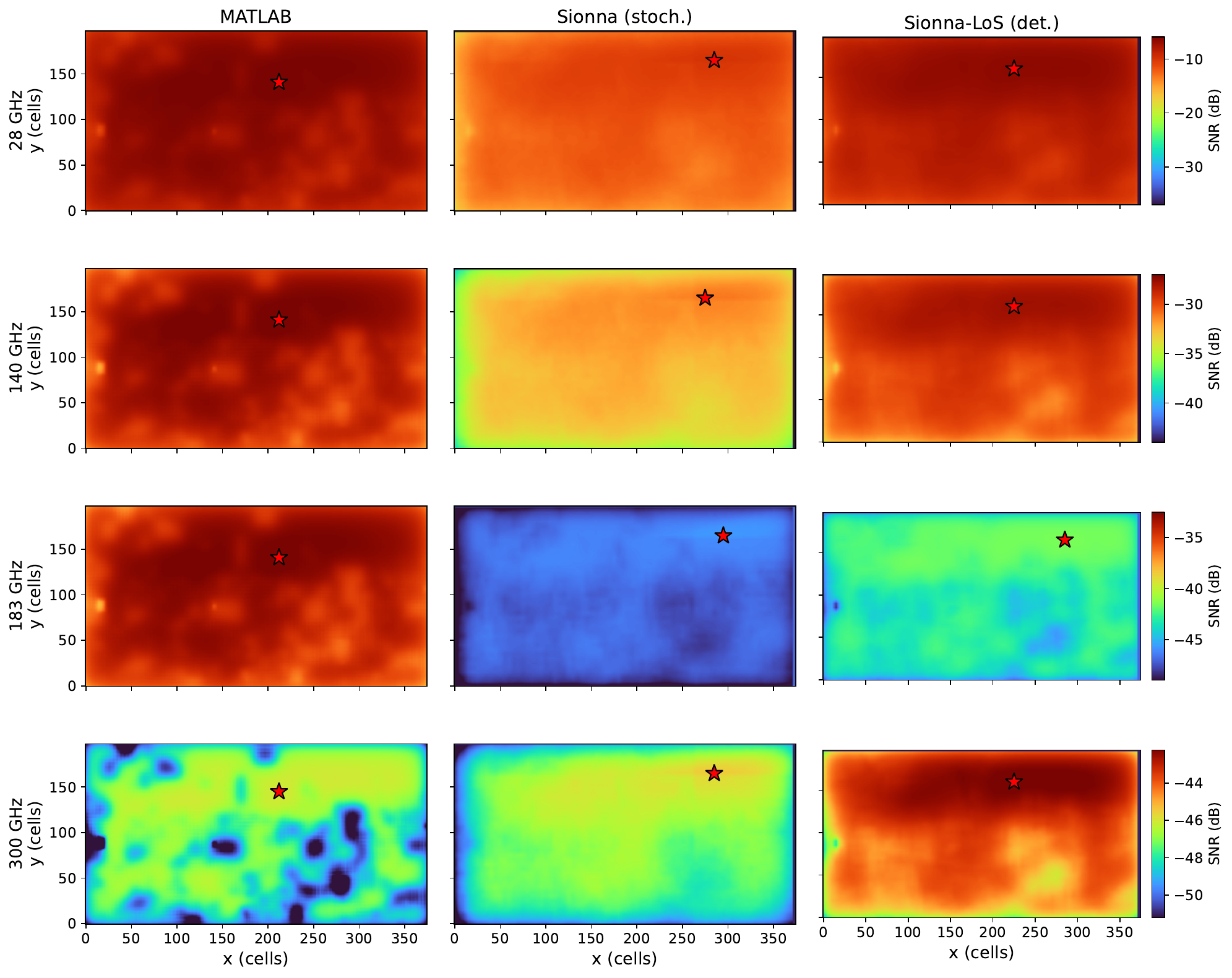}
\caption{Spatial SNR maps over $\mathcal{P}_z$ at $z = \SI{100}{\meter}$.
Red stars denote per-pipeline argmaxes.}
\label{fig:heatmaps}
\end{figure}

Fig.~\ref{fig:cdfs} overlays the empirical CDFs of
$\Phi^{\mathrm{M}}, \Phi^{\mathrm{S}}, \Phi^{\mathrm{L}}$ at
$z = \SI{100}{\meter}$ across the four carriers, and Fig.~\ref{fig:heatmaps}
shows the corresponding spatial maps with per-pipeline argmax markers. We
highlight three carrier-specific phenomena:
\begin{itemize}
  \item At $28$ and $140$~\unit{\giga\hertz}, $\Phi^{\mathrm{S}}$ lies
        $\approx 5$--$6$~\unit{\dB} below $\Phi^{\mathrm{M}}$, but
        $\Phi^{\mathrm{L}}$ is only $\approx 1.5$--$1.7$~\unit{\dB} below:
        most of the apparent gap is sampling-induced.
  \item At $183$~\unit{\giga\hertz}, both $\Phi^{\mathrm{S}}$ and
        $\Phi^{\mathrm{L}}$ are heavily shifted relative to
        $\Phi^{\mathrm{M}}$ (by $-13.2$ and $-9.2$~\unit{\dB},
        respectively); the JS divergence saturates at $\ln 2 \approx 0.693$
        in both cases, indicating practically disjoint distributions.
  \item At $300$~\unit{\giga\hertz}, $\Phi^{\mathrm{S}}$ is nearly aligned
        with $\Phi^{\mathrm{M}}$ (bias $+0.5$~\unit{\dB},
        $D_{\mathrm{JS}} = 0.046$). The deterministic-LoS pipeline,
        however, exposes a $+3.8$~\unit{\dB} structural offset
        ($\Phi^{\mathrm{L}}$ above $\Phi^{\mathrm{M}}$) and
        $D_{\mathrm{JS}} = 0.62$. The stochastic agreement at this carrier
        is therefore coincidental rather than physical.
\end{itemize}


\subsection{Reality-Gap Quantification}\label{sec:res-gap}

\begin{figure}[t]
\centering
\includegraphics[width=0.86\linewidth]{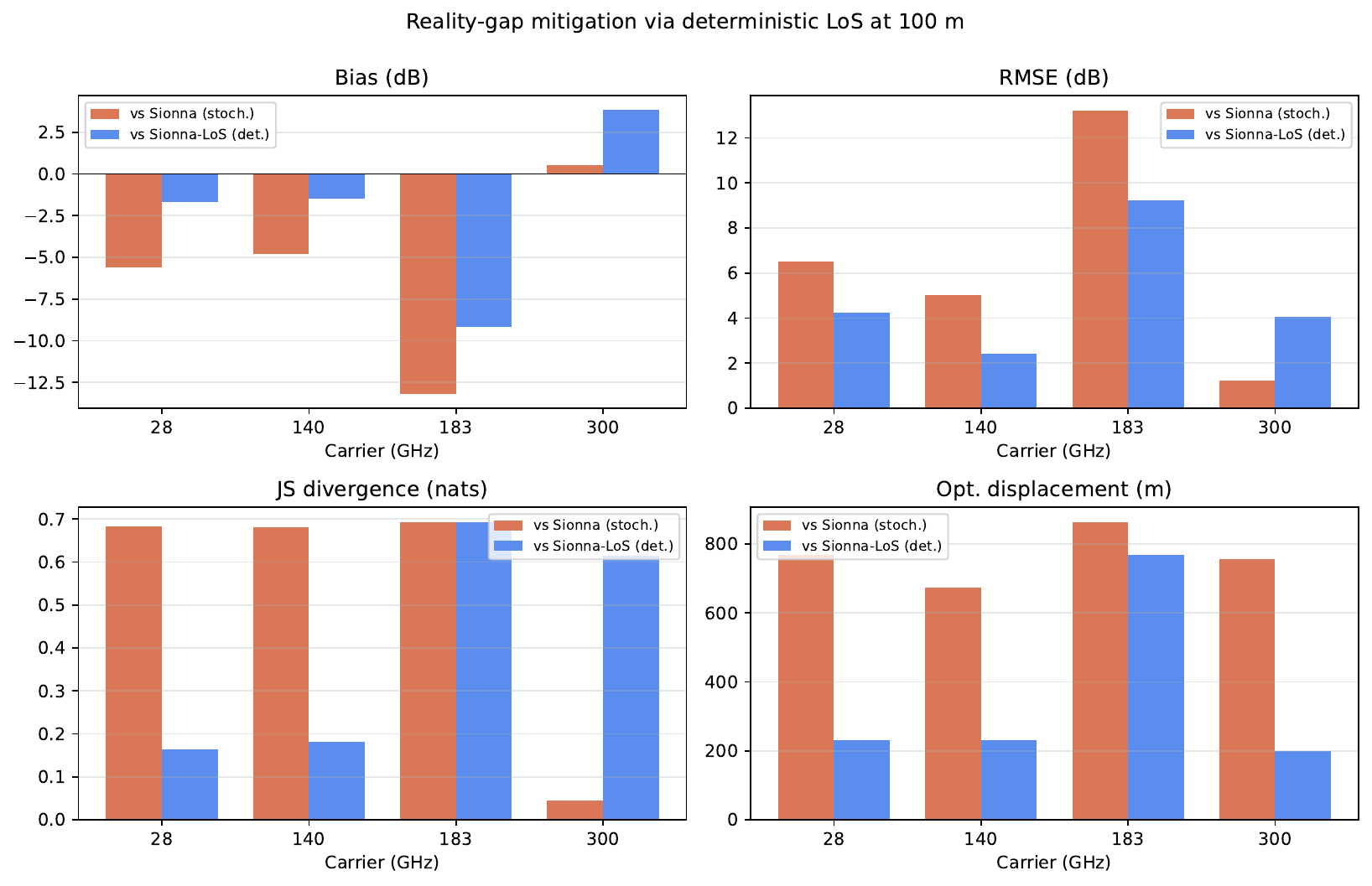}
\caption{Reality-gap metrics at $z = \SI{100}{\meter}$ grouped by target
pipeline.}
\label{fig:gapbars}
\end{figure}

Table~\ref{tab:gap} reports the four reality-gap metrics
\eqref{eq:bias}--\eqref{eq:disp} and the regret ratio \eqref{eq:regret} at
$z = \SI{100}{\meter}$ for both ray-traced targets, and
Fig.~\ref{fig:gapbars} gives the corresponding side-by-side bar chart.
Three regimes emerge.

\paragraph{Sampling-dominated regime ($28$, $140$~\unit{\giga\hertz})}
The bias against stochastic Sionna is
$\mathrm{bias}^{\mathrm{S}} \in \{-5.58, -4.80\}$~\unit{\dB}; mitigation
reduces this to $\mathrm{bias}^{\mathrm{L}} \in \{-1.69, -1.47\}$~\unit{\dB}.
The decomposition \eqref{eq:decomp} attributes
$\sim\SI{3.9}{\dB}$ at $28$~\unit{\giga\hertz} (resp.\ $\sim\SI{3.3}{\dB}$
at $140$~\unit{\giga\hertz}) to Monte-Carlo undersampling and the residual
$\sim\SI{1.5}{\dB}$ to genuine geometric/material differences. The JS
divergence collapses from $0.683$/$0.681$ to $0.164$/$0.181$, indicating
near-overlapping distributions. The optimum-deployment displacement
contracts dramatically, from $\SI{768}{\meter}$ to $\SI{230}{\meter}$ at
$28$~\unit{\giga\hertz} ($-70\%$), and the regret improves from
$0.850$ to $0.956$.

\paragraph{Absorption-dominated regime ($183$~\unit{\giga\hertz})}
The $\SI{183.31}{\giga\hertz}$ water-vapour resonance is represented
in the analytical pipeline, it appears as the second Lorentzian
($\tilde{\nu}_2 = 6.11~\mathrm{cm}^{-1}$) of the
Kokkoniemi--Lehtom\"aki fit \cite{kokkoniemi2021los}, but a single
Lorentzian under-predicts the peak amplitude and wing tails of the full
ITU-R~P.676-13 line tables, and this discrepancy compounds over the
multi-cell aggregate of \eqref{eq:plavg}. Mitigation reduces
$\mathrm{bias}^{\mathrm{S}} = -13.2$~\unit{\dB} only to
$\mathrm{bias}^{\mathrm{L}} = -9.2$~\unit{\dB}, and $D_{\mathrm{JS}}$
remains saturated at $\ln 2$ in both cases: the two SNR distributions are
practically disjoint. The displacement contracts modestly
($\SI{864}{\meter} \!\to\! \SI{768}{\meter}$).

\paragraph{Coincidental-agreement regime ($300$~\unit{\giga\hertz})}
Stochastic Sionna agrees with the analytical model
($\mathrm{bias}^{\mathrm{S}} = +0.50$~\unit{\dB},
$D_{\mathrm{JS}}^{\mathrm{S}} = 0.046$, $\rho^{\mathrm{S}} = 0.887$),
suggesting at first sight that the analytical channel is sufficient. The
deterministic-LoS pipeline, however, reveals
$\mathrm{bias}^{\mathrm{L}} = +3.84$~\unit{\dB} and
$D_{\mathrm{JS}}^{\mathrm{L}} = 0.615$. By
\eqref{eq:decomp}, the apparent $+0.5$~\unit{\dB} bias decomposes into a
$-3.3$~\unit{\dB} sampling component and a $+3.8$~\unit{\dB} physics
component that almost exactly cancel. Inspecting the spatial maps
(Fig.~\ref{fig:heatmaps}, bottom row) confirms the mechanism: the
$\mathcal{M}$ map exhibits high-spatial-frequency speckle absent from both
ray-traced pipelines, traceable to the position-dependent absorption
polynomial; the smooth $\mathcal{L}$ map is dominated by the
geometric-LoS gradient; and stochastic Sionna averages the two effects
into accidental alignment.

\begin{table}[t]
\caption{Reality-gap metrics at $z=\SI{100}{m}$ for the two ray-traced
targets.}
\label{tab:gap}
\centering\small
\setlength{\tabcolsep}{3.5pt}
\renewcommand{\arraystretch}{1.05}
\begin{tabular}{l c r r r r r}
\toprule
Target & Carrier & Bias~(\unit{\dB}) & RMSE~(\unit{\dB}) & $D_{\mathrm{JS}}$ & $\delta$~(m) & $\rho$ \\
\midrule
\multirow{4}{*}{$\mathcal{S}$ stoch.}
& $28$~GHz   & $-5.58$  & $6.51$   & $0.683$ & $768$ & $0.850$ \\
& $140$~GHz  & $-4.80$  & $5.02$   & $0.681$ & $674$ & $0.885$ \\
& $183$~GHz  & $-13.20$ & $13.22$  & $0.693$ & $864$ & $0.880$ \\
& $300$~GHz  & $+0.50$  & $1.25$   & $0.046$ & $757$ & $0.887$ \\
\midrule
\multirow{4}{*}{$\mathcal{L}$ det.}
& $28$~GHz   & $-1.69$  & $4.26$   & $0.164$ & $230$ & $0.956$ \\
& $140$~GHz  & $-1.47$  & $2.44$   & $0.181$ & $230$ & $0.954$ \\
& $183$~GHz  & $-9.19$  & $9.22$   & $0.693$ & $768$ & $0.957$ \\
& $300$~GHz  & $+3.84$  & $4.05$   & $0.615$ & $198$ & $0.967$ \\
\bottomrule
\end{tabular}
\end{table}

\subsection{Implications for RL Policy Transfer}\label{sec:res-transfer}

The optimum-displacement and regret columns of Table~\ref{tab:gap} are the
operational quantities for sim-to-real transfer. Although $\delta$ ranges
from $\SI{198}{\meter}$ to $\SI{864}{\meter}$, the linear-domain regret
$\rho^{\mathrm{L}} \geq 0.93$ at every carrier, so the analytical-optimal
deployment never costs more than $\sim 7\%$ in linear SNR relative to the
deterministic ray-traced optimum. Operationally, $\rho$ bounds the SNR
sacrificed by deploying at the analytical optimum ($\leq\SI{0.3}{\dB}$
here), while $\delta$ shows the optimum's \emph{location} transfers far
less reliably than its \emph{value}. Three operational conclusions
follow.
At $28$/$140$~\unit{\giga\hertz} the analytical channel is sufficient for
placement: residual bias is within link-budget margin ($\sim\SI{1.5}{\dB}$)
and regret is $\geq 0.95$. At $183$~\unit{\giga\hertz} an absorption-grade
upgrade (Kokkoniemi $\to$ ITU-R~P.676) is necessary; LoS modelling alone
does not close the gap, and $D_{\mathrm{JS}}$ saturation indicates that
the analytical and ray-traced SNR distributions are operationally
inequivalent. At $300$~\unit{\giga\hertz}, although the deterministic-LoS
comparison reveals a $+3.8$~\unit{\dB} structural offset, the regret ratio
remains $0.967$: the analytical model is still adequate for placement
decisions despite being structurally inequivalent in the absolute SNR sense.

\section{Conclusion}\label{sec:conclusion}

We presented a quantitative reality-gap analysis for an RL-controlled UAV
relay across four carriers spanning mmWave to sub-THz, using a real urban
map and a publicly reproducible Sionna pipeline. The disagreement between
a simplified geometric channel and ray tracing is sharply carrier
dependent: at $28$/$140$~\unit{\giga\hertz} the apparent bias is dominated
by Monte-Carlo undersampling and is removed by deterministic-LoS
recomputation; at $183$~\unit{\giga\hertz} a residual bias persists,
tracing to the absorption-line model; and at $300$~\unit{\giga\hertz} the
stochastic ray tracer coincidentally agrees with the analytical model
while the deterministic-LoS pipeline exposes a structural offset. The
optimum-deployment displacement is on the order of hundreds of metres but
the linear-domain regret ratio remains $\geq 0.93$ across all carriers,
suggesting that policies trained on simplified channels remain useful but
should be fine-tuned on ray tracing for line-adjacent carriers. The study is
limited to a single city, single-seed stochastic runs, static single-UAV
placement with isotropic antennas, a fairness-agnostic mean-gain
objective, and $D{=}1$ specular ray tracing without diffraction.
Future work
will extend the framework to multi-UAV, realistic user distribution settings and to differentiable
fine-tuning that uses the analytical model as a prior and ray tracing as the high-fidelity oracle.

\section*{Acknowledgment}
{\footnotesize
This publication was made possible in part by by NPRP14C-0909-210008 from the Qatar Research, Development and Innovation (QRDI) Fund (a member of Qatar Foundation) and by research funding from Hamad Bin Khalifa University under the Thematic Research Grant Program Cycle 3, in part by the U.S. Department of Energy, Office of Science, Office of Advanced Scientific Computing Research, Early Career Research Program under Award Number DE-SC0023957, and in part by the National Science Foundation under Grant No. 2549088.\par}

\bibliographystyle{IEEEtran}
\bibliography{globecom2026}

\end{document}